\documentclass[12pt]{article}
\usepackage{latexsym,amsfonts,amssymb}

\newtheorem{theorem}{Theorem}
\newtheorem{definition}{Definition}
\newtheorem{lemma}{Lemma}
\newtheorem{corollary}{Corollary}
\newtheorem{proposition}{Proposition}

\begin{document}

\title{On separable Pauli equations}

\author{Alexander Zhalij\thanks{e-mail: zhaliy@imath.kiev.ua} \\ \small
Institute of Mathematics of the Academy of Sciences of Ukraine,\\
\small Tereshchenkivska Street 3, 01601 Kyiv-4, Ukraine}

\date{}

\maketitle


\begin{abstract}
We classify (1+3)-dimensional Pauli equations for a
spin-$\frac{1}{2}$ particle interacting with the electro-magnetic
field, that are solvable by the method of separation of variables.
As a result, we obtain the eleven classes of vector-potentials of
the electro-magnetic field $A(t, \vec x)=(A_0(t, \vec x)$, $\vec
A(t, \vec x))$ providing separability of the corresponding Pauli
equations. It is established, in particular, that the necessary
condition for the Pauli equation to be separable into second-order
matrix ordinary differential equations is its equivalence to the
system of two uncoupled Schr\"odinger equations. In addition, the
magnetic field has to be independent of spatial variables. We
prove that coordinate systems and the vector-potentials of the
electro-magnetic field providing the separability of the
corresponding Pauli equations coincide with those for the
Schr\"odinger equations. Furthermore, an efficient algorithm for
constructing all coordinate systems providing the separability of
Pauli equation with a fixed vector-potential of the
electro-magnetic field is developed. Finally, we describe all
vector-potentials $A(t, \vec x)$ that (a) provide the separability
of Pauli equation, (b) satisfy vacuum Maxwell equations without
currents, and (c) describe non-zero magnetic field.
\end{abstract}

\section{Introduction}
A quantum mechanical system consisting of a spin--${1\over 2}$
charged particle, moving with momentum $\vec p$ \ in a
time-dependent electro-magnetic field with the four-component
vector-potential $(A_0, \vec A)$, is described in a
non-relativistic approximation by the Pauli equation (see, e.g.,
\cite{lanIV})
\begin{equation}
\label{1.1} \left(p_0 - e A_0(t,{\vec x})- \left(\vec p - e \vec
A(t, {\vec x})\right)^2+e\vec\sigma\vec H\right)\psi(t,\vec x)=0.
\end{equation}
Here $\psi (t,\vec x)$ is the two-component wave function in three
space dimensions $\vec x=(x_1,x_2,x_3)$, $\vec H={\rm rot}\ \vec
A$ is the magnetic field, and $\vec\sigma =(\sigma_1, \sigma_2,
\sigma_3)$ is a vector consisting of three Pauli matrices
\begin{equation}
\label{sigma} \sigma_1=\left(\begin{array}{cc} 0 & 1\\1 & 0
\end{array}\right),\quad
\sigma_2=\left(\begin{array}{cc} 0 & -i\\i & 0
\end{array}\right),\quad
\sigma_3=\left(\begin{array}{cc} 1 & 0\\0 & -1
\end{array}\right).
\end{equation}
Hereafter we use the notations
\begin{equation}
\label{p} p_0=i{\partial\over\partial t},\quad \vec p=-
i{\vec\nabla},\quad a=1,2,3,
\end{equation}
and summation over the repeated Latin indices from 1 to 3 is
implied.

As the Pauli equation has variable coefficients, we cannot apply
the standard Fourier transformation. The only regular way for
solving (\ref{1.1}) is the classical method of separation of
variables in curvilinear coordinate systems. In this respect, a
natural question arises, which equations of the form (\ref{1.1})
are separable, namely, which potentials $A_0, \vec A$ allow for
separability of the Pauli equation in some curvilinear coordinate
system? One of the principal objectives of the present article is
to provide an efficient algorithm for answering these kinds of
questions for systems of partial differential equations. It is
essentially based on the results of the paper \cite{zhd99b}, where
the problem of separation of variables in the Schr\"odinger
equation has been analyzed in detail. As the Pauli equation
(\ref{1.1}) differs from the Schr\"odinger equation by the term
$e\vec\sigma\vec H$ only, it is natural to attempt modifying the
technique employed in \cite{zhd99b} in order to make it applicable
to system of partial differential equations (PDEs) (\ref{1.1}).

Integrable Hamiltonian systems with velocity--dependent potentials
have been studied for the case $n=2$, i.e., in a Euclidean plane
by Winternitz with co-authors \cite{dor85,mcs00}. Recently Benenti
with co-authors \cite{ben01} studied  the problem of separation of
variables in the stationary Hamilton-Jacobi equation with
vector-potential from a geometrical point of view.

The problem of separation of variables for linear systems of
first-order partial differential equations such as the Dirac
equation has been repeatedly addressed by Shapovalov and Bagrov
with co-authors \cite{sha72,bag90} and by Kalnins and Miller with
co-authors \cite{kal86,mill88,mill89}. They developed a symmetry
approach to the separation of variables in the Dirac equation
where separability is characterized by the existence of a complete
set of first-order matrix symmetry operators.

Symmetry and supersymmetry properties of the Pauli equations are
studied in \cite{tka98,nik99,nie99}. Let us also mention the paper
\cite{kos93}, where physical aspects of the problem of separation
of variables in some (1+3)-dimensional Pauli equations with time
dependent potentials are studied, and the monograph \cite{bag90},
where some classes of exact solutions of the Pauli equation are
presented.

With all the variety of approaches to separation of variables in
PDEs one can notice the three generic principles, namely,
\begin{enumerate}
\item[a)]{Representation of a solution to be found in a separated
(factorized) form via several functions of one variable.}
\item[b)]{Requirement that the above mentioned functions of
one variable should satisfy some ordinary differential
equations.}
\item[c)]{Dependence of so found solution on several arbitrary
(continuous or discrete) parameters, called spectral parameters
or separation constants.}
\end{enumerate}
By a proper formalizing of the above features we have formulated
in \cite{zhd99b} an algorithm for variable separation in the
Schr\"odinger equation with vector-potential. Below we generalize
this algorithm for the case of system of PDEs (\ref{1.1}).

To have a right to talk about description of {\it all} potentials
and {\it all} coordinate systems enabling us to separate the Pauli
equation, one needs to provide a rigorous definition of separation
of variables. The definition we intend to use is based on ideas
contained in the paper by Koornwinder \cite{koo80}.

Let us introduce a new coordinate system $t$,
$\omega_a=\omega_a(t,{\vec x})$, $a=1,2,3$, where $\omega_a$ are
real-valued functions, functionally independent with respect to
the spatial variables $x_1$, $x_2$, $x_3$, i.e.:
\begin{equation}
\label{3a} {\rm det}\left\|{\partial\omega_a\over \partial
x_b}\right\|_{a,b=1}^3\ne 0.
\end{equation}

For a solution to be found we adopt the following separation Ansatz:
\begin{equation}
\label{2.1} \psi(t,{\vec x})=Q(t,{\vec
x})\varphi_0(t)\prod\limits_{a=1}^3\, \varphi_a\left(\omega_a(t,
{\vec x}), \vec \lambda\right)\chi,
\end{equation}
where $Q,\ \varphi_\mu,\ (\mu=0,1,2,3)$ are non-singular $2\times
2$--matrix functions of the indicated variables and $\chi$ is an
arbitrary two-component constant column. What is more, the
condition of commutativity of the matrices $\varphi_\mu$ is
imposed, namely,
\begin{equation}
\label{2.4} [\varphi_\mu,\
\varphi_\nu]=\varphi_\mu\varphi_\nu-\varphi_\nu\varphi_\mu=0,\quad
\mu,\nu=0,1,2,3.
\end{equation}
Note that the restriction (\ref{2.4}) is an extra requirement,
which narrows the class of separable Pauli equations. However,
without this condition an efficient handling of the Ans\"atze of
the form (\ref{2.4}) seems to be impossible. At least, in all
papers devoted to variable separation in a systems of PDEs the
condition of commutativity is imposed (explicitly or implicitly).

\begin{definition}
\label{oz1} We say that the Pauli equation (\ref{1.1}) admits
separation of variables in a coordinate system $t$,
$\omega_a=\omega_a(t,\vec x)$, $a=1,2,3$, if there are
non--singular $2\times 2$--matrix function $Q(t,{\vec x})$ and
four matrix ordinary differential equations
\begin{equation}
\label{2.2}
\begin{array}{l}
i\dot\varphi_0=-\left(P_{00}(t)+P_{0b}(t)\lambda_b\right)\varphi_0,\\[2mm]
\ddot\varphi_a=\left(P_{a0}(\omega_a) +
P_{ab}(\omega_a)\lambda_b\right)\varphi_a,\quad a=1,2,3,
\end{array}
\end{equation}
jointly depending in an analytic way on three independent complex
parameters $\lambda_1,\lambda_2,\lambda_3$ (separation constants),
such that, for each triplet $(\lambda_1,\lambda_2,\lambda_3)$ and
for each set of solutions $\varphi_0(t)$, $\varphi_1(\omega_1)$,
$\varphi_2(\omega_2)$, $\varphi_3(\omega_3)$ of (\ref{2.2}),
function (\ref{2.1}) under condition (\ref{2.4}) is a solution of
(\ref{1.1}).
\end{definition}

In the above formulas $P_{\mu\nu}, \mu,\nu=0,1,2,3$ are some
complex $2\times 2$--matrix functions of the indicated variables.

\begin{definition}
Three complex parameters $\lambda_1,\lambda_2,\lambda_3$
in (\ref{2.2}) are called independent, if the equality
\begin{equation}
\label{2.3}
{\rm rank}\, \left \|P_{\mu a}\right\|_{\mu =0\; a=1}^{3\quad\; 3}=6.
\end{equation}
holds, whenever
$\varphi_0(t)\varphi_1(\omega_1)\varphi_2(\omega_2)\varphi_3(\omega_3)\ne
0$.
\end{definition}

Condition (\ref{2.3}) secures essential dependence of a
solution with separated variables on the separation constants
$\vec \lambda$.

Note, that putting $Q=I, \omega_a=x_a, a=1,2,3$ in (\ref{2.4})
yields the standard separation of variables in the Cartesian
coordinate system. Next, choosing the spherical coordinates as
$\omega_1, \omega_2, \omega_3$ we arrive at the variable
separation in the spherical coordinate system and so on. The
principal task is describing all possible forms of the functions
$Q, \omega_a, a=1,2,3$, that provide separability of the Pauli
equation in the sense of the definition given above. Solution of
this problem, in its turn, requires describing the functions
$A_0,\ldots, A_3$ that enable variable separation in the Pauli
equation in the corresponding coordinate system. More precisely,
we will need to solve the two mutually connected principal
problems:
\begin{itemize}
\item[{\bf --}]
to describe {\it all} cases of coefficients, for which the
corresponding Pauli equation  (\ref{1.1}) is separable (in the
sense of definition 1) in at least one coordinate system;
\item[{\bf --}]
to construct {\it all} coordinate systems that allow for
separation of variables (in the sense of definition 1) in the
Pauli equation (\ref{1.1}) with some fixed vector-potential $(A_0,
\vec A)$.
\end{itemize}

Note, that formulas (\ref{2.1})--(\ref{2.3}) form the input data
of the method. We can change these conditions and thereby modify
the definition of separation of variables. For instance, we can
change the order of the reduced equations (\ref{2.2}) or the
number of essential parameters $\lambda_a$ (a more detailed
analysis of this problem for the Schr\"odinger equation can be
found in \cite{zhd97}). So, our claim of obtaining the {\it
complete description} of vector-potentials and coordinate systems
providing separation of variables in (\ref{1.1}) makes sense only
within the framework of definition 1. If one uses a more general
definition, it might be possible to construct new coordinate
systems and vector--potentials providing separability of equation
(\ref{1.1}). But all solutions of the Pauli equation with
separated variables known to us fit into the above suggested
scheme.

Transformations
\begin{equation}
\label{la.1}
\lambda_a\to\lambda_a'=\Lambda_a(\lambda_1,\lambda_2,\lambda_3),
\quad a=1,2,3
\end{equation}
under condition
\begin{equation}
\label{la.2} {\rm det}\left\|{\partial\Lambda_a\over \partial
\lambda_b}\right\|_{a,b=1}^3\ne 0.
\end{equation}
preserve the form of relations (\ref{2.1})--(\ref{2.3}). So we can
regard the corresponding spectral parameters $\vec \lambda$ and
$\vec \lambda'$ as {\it equivalent} ones. Within the framework of
this equivalence relation we can choose $\vec \lambda$ in such a
way that all matrices $P_{\mu\nu}, \mu,\nu=0,1,2,3$ in reduced
equations (\ref {2.2}) are Hermitian ones and parameters
$\lambda_1, \lambda_2, \lambda_3$ are real numbers.

Next, we introduce an equivalence relation on the set of all
vector--potentials $A_0(t, \vec x)$, $\vec A(t, \vec x)$
providing separability of equation (\ref{1.1}), on the sets
of solutions with separated variables and corresponding
coordinate systems.
\begin{definition}
We say that two vector--potentials $A(t, \vec x)$ and
$A'(t, \vec x)$ are equivalent if they are
transformed one into another by the gauge transformation
\begin{equation}
\vec A\to \vec A'=\vec A+\vec\nabla f,\quad A_0\to
A'_0=A_0-{\partial f\over \partial t},\label{calib2}
\end{equation}
where $f=f(t,\vec x)$ is an arbitrary smooth function.
\end{definition}
For the Pauli equations to be invariant with respect to the above
transformation, the wave function $\psi(t,\vec x)$ is to be
transformed according to the rule
\begin{equation}
\label{calib3}
\psi\to \psi'=\psi \exp (ief)
\end{equation}
Indeed, if the transformations (\ref{calib2})--(\ref{calib3}) in
the Pauli equation (\ref{1.1}) are performed one after another, we
obtain the initial equation, provided we replace the functions $\vec A, A_0, \psi$
with $\vec A', A'_0, \psi'$.

Note that the system of PDEs (\ref{1.1}) admits a wider
equivalence group from the point of view of the standard theory of
partial differential equations (Shapovalov and Sukhomlin
\cite{sha74}). However, this group cannot be regarded as an
equivalence group within the context of quantum mechanics, since
allowed transformations of the wave function must preserve the
probability density $\psi^*\psi$. And it is straightforward to
check the the wider Shapovalov and Sukhomlin equivalence group
breaks this rule, because it, generally speaking, does not
preserve $\psi^*\psi$. By this very reason, we restrict our
considerations to the gauge transformations only.

\begin{definition}
Two solutions of the Pauli equation with separated variables are
called equivalent if they can be transformed one into another by
group transformations from the Lie transformation group admitted
by Pauli equation (\ref{1.1}). Moreover, solutions of the Pauli
equation with separated variables having equivalent (in the sense
of equivalence relation (\ref{la.1})-(\ref{la.2})) spectral
parameters $\vec\lambda$ are equivalent.
\end{definition}
\begin{definition}
Two coordinate systems $t$, $\omega_1$, $\omega_2$, $\omega_3$ and
$t'$, $\omega_1'$, $\omega_2'$, $\omega_3'$ are called equivalent
if they give equivalent solutions with separated variables. In
particular, two coordinate systems are equivalent if the
corresponding Ans\"atze (\ref{2.1}) are transformed one into
another by reversible transformations of the form
\begin{eqnarray}
&&t\to t'=f_0(t),\quad \omega_a\to\omega_a'=f_a(\omega_a),\quad
a=1,2,3,\label{1.equ}\\ &&Q\to Q'=Ql_0(t) l_1(\omega_1)
l_2(\omega_2)l_3(\omega_3),\label{0.6b}
\end{eqnarray}
where $f_0,\ldots, f_3$ are some smooth functions and
$l_0,\ldots, l_3$ are some smooth $2\times
2$--matrix functions of the indicated variables.
\end{definition}
Indeed, transformations (\ref{1.equ}) and (\ref{0.6b}) preserve
the form of Ans\"atze (\ref{2.1}). So after completing the procedure of
separation of variables in these coordinate systems we obtain the
same solutions with separated variables.

These equivalence relations reflect the freedom in choice of the
functions $Q, \omega_1, \omega_2, \omega_3$ and separation
constants $\lambda_1, \lambda_2, \lambda_3$ preserving the form of
the conditions (\ref{2.1})--(\ref{2.3}). They split the set of all
possible vector--potentials, providing separability of equation
(\ref{1.1}), and sets of solutions with separated variables and
corresponding coordinate systems into equivalence classes. In a
sequel, when presenting the corresponding lists we will give only
one representative for each equivalence class.

\section{Classification of separable Pauli equations (\ref{1.1})}

In this section we obtain an exhaustive classification of the
Pauli equations solvable within the framework of the approach
described in the Introduction. Furthermore, we describe
curvilinear coordinate systems enabling separation of variables in
(\ref{1.1}).

Using the equalities (\ref{2.2}) and (\ref{2.4}) we get
\[
[P_{\mu 0}+P_{\mu a}\lambda_a, P_{\nu 0}+P_{\nu a}\lambda_a]=0.
\]
Splitting the expression with respect to $\lambda_a$ yields
\begin{equation}
[P_{\mu\alpha}, P_{\nu\beta}]+[P_{\mu\beta}, P_{\nu\alpha}]=0,
\label{2.5}
\end{equation}
where $\mu,\nu,\alpha,\beta=0,1,2,3$ and henceforth summation over
repeated Greek indices is not used. Choosing $\alpha=\beta$ we
have
\[
[P_{\mu\alpha}, P_{\nu\alpha}]=0.
\]
Taking into account this equality and the fact that any Hermitian
$(2\times 2)$--matrix can be represented as a linear combination
of the unit and Pauli matrices (\ref{sigma}), we get the following
form of $P_{\mu\alpha}$:
\begin{equation}
\label{2.6}
P_{\mu\alpha}=F_{\mu\alpha}(\omega_\mu)I+G_{\mu\alpha}(\omega_\mu)\vec
s_\alpha\vec\sigma,
\end{equation}
where $F_{\mu\alpha}, G_{\mu\alpha}$ are some smooth scalar
functions of the indicated variables, $\omega_0=t$ and $\vec
s_\alpha$ is a constant three-component vector. Substitution of
expression (\ref{2.6}) into (\ref{2.5}) yields
\[
(G_{\mu\alpha}G_{\nu\beta}-G_{\mu\beta}G_{\nu\alpha})[\vec
s_\alpha\vec\sigma, \vec s_\beta\vec\sigma]=0.
\]
From this equality we conclude that there are two distinct cases:
either $\vec s_\alpha\sim\vec s_\beta$ or $G_{\mu\alpha}\sim
G_{\mu\beta}$. In view of this fact we get the two possible forms
for the equations (\ref{2.2}):
\begin{eqnarray}
&&i\dot\varphi_0=-\left(F_{00}(t)+F_{0b}(t)\lambda_b+
(G_{00}(t)+G_{0b}(t)\lambda_b)\vec
s\vec\sigma\right)\varphi_0,\label{2.7a}\\
&&\ddot\varphi_a=\left(F_{a0}(\omega_a) +
F_{ab}(\omega_a)\lambda_b+ (G_{a0}(\omega_a)
+G_{ab}(\omega_a)\lambda_b)\vec s\vec\sigma\right)\varphi_a
\nonumber
\end{eqnarray}
and
\begin{eqnarray}
&&i\dot\varphi_0=-\left(F_{00}(t)+F_{0b}(t)\lambda_b+ G_0(t)(\vec
s_0+ \vec s_b\lambda_b)\vec\sigma\right)\varphi_0,\label{2.8a}\\
&&\ddot\varphi_a=\left(F_{a0}(\omega_a) +
F_{ab}(\omega_a)\lambda_b+ G_a(\omega_a)(\vec s_0+ \vec
s_b\lambda_b)\vec\sigma\right)\varphi_a,\nonumber
\end{eqnarray}
with $a=1,2,3$.

Definition 1 is quite algorithmic in the sense that it contains a
regular algorithm of variable separation in Pauli equation
(\ref{1.1}). Formulas (\ref{2.1}), (\ref{2.7a})--(\ref{2.8a}) form
the input data of the method. The principal steps of the procedure
of variable separation in Pauli equation (\ref{1.1}) are as
follows.
\begin{enumerate}
\item
We insert the Ansatz (\ref{2.1}) into the Pauli equation and
express the derivatives $\dot\varphi_0$, $\ddot\varphi_1$,
$\ddot\varphi_2$, $\ddot\varphi_3$ in terms of functions
$\varphi_0,\, \varphi_1,\, \varphi_2,\, \varphi_3$, using
equations (\ref{2.7a})--(\ref{2.8a}).
\item
We regard $\varphi_0,\, \varphi_1,\, \varphi_2,\, \varphi_3,\,
\lambda_1,\, \lambda_2,\, \lambda_3$ as new independent variables
$y_1$, $\ldots$, $y_{7}$. As the functions $Q$, $\omega_1$,
$\omega_2$, $\omega_3$, $A_0$, $A_1$, $A_2$, $A_3$ are independent
on the variables $y_1,\ldots, y_{7}$, we can demand that the
obtained equality is transformed into identity under arbitrary
$y_1$, $\ldots$, $y_{7}$. In other words, we should split the
equality with respect to these variables under condition of
commutativity (\ref{2.4}). After splitting we get an
overdetermined system of nonlinear partial differential equations
for unknown functions $Q$, $\omega_1$, $\omega_2$, $\omega_3$,
$A_0$, $A_1$, $A_2$, $A_3$.
\item
After solving the above system we get an exhaustive description
of vector--potentials $A(t,\vec x)$ providing separability
of the Pauli equation and corresponding coordinate systems.
\end{enumerate}
Having performed the first two steps of the above algorithm we
obtain the system of nonlinear matrix PDEs:
\begin{eqnarray*}
&(i)\ &
  {\partial \omega_b\over \partial x_a}{\partial \omega_c\over \partial x_a}=0,\quad
  b\not = c,\quad b,c=1,2,3;\\
&(ii)\ &
  \sum\limits_{a=1}^{3}\,
  F_{ab}(\omega_a) {\partial \omega_a\over \partial x_c}{\partial \omega_a\over \partial x_c}
  = F_{0b}(t),\quad b=1,2,3;\\
&(iiia)\ &
  \mbox{For case of reduced equations given by (\ref{2.7a})}\\
&&\sum\limits_{a=1}^{3}\,
  G_{a\mu}(\omega_a) {\partial \omega_a\over \partial x_c}{\partial \omega_a\over \partial x_c} =
  G_{0\mu}(t),\quad \mu=0,1,2,3;\\
&(iiib)\ &
  \mbox{For case of reduced equations given by (\ref{2.8a})}\\
&&\sum\limits_{a=1}^{3}\,
  G_a(\omega_a) {\partial \omega_a\over \partial x_c}{\partial\omega_a\over \partial x_c} = G_0(t),\\
&(iv)\ &
  2\left({\partial Q \over \partial x_b}-i e Q A_b\right){\partial \omega_a\over \partial x_b} +
  Q\left(i{\partial\omega_a\over\partial t} + \Delta \omega_a\right) = 0,\quad a=1,2,3;\\
&(v)\ &
  Q \sum\limits_{a=1}^{3}\, F_{a0}(\omega_a) {\partial \omega_a\over \partial
  x_b}{\partial \omega_a\over \partial x_b} + i{\partial Q\over \partial t} + \Delta Q -
  2 i e A_b{\partial Q \over \partial x_b}+\\
&&\phantom{Q \sum\limits_{i=1}^{3}}
  + \left(-F_{00}(t) - i e {\partial A_b\over \partial x_b} - e A_0 -
  e^2 A_bA_b + e\vec\sigma\vec H\right)Q=0.
\end{eqnarray*}
Thus the problem of variable separation in the Pauli equation
reduces to integrating of a system of nonlinear PDEs for eight
unknown functions $A_0, A_1, A_2, A_3, Q, \omega_1, \omega_2,
\omega_3$ of four variables $t, \vec x$. What is more, some
coefficients are arbitrary matrix functions which should be
determined in the process of integrating of the system of PDEs
$(i)-(v)$. We succeeded in constructing the general solution of
the latter which yields, in particular, all possible
vector-potentials $A(t, \vec x)=(A_0(t, \vec x),\ldots, A_3(t,
\vec x))$ such that Pauli equation (\ref{1.1}) is solvable by the
method of separation of variables.

In view of (\ref{2.3}) we can always choose from each set of the
equations $(ii)-(iiia)$ and $(ii)-(iiib)$ three such equations
that the matrix of coefficients of $\omega_{ax_c} \omega_{ax_c}$
(a=1,2,3) is non-singular. It is called the St\"ackel matrix
\cite{sta91}. The system consisting of these three equations and
of the equations $(i)$ was integrated in \cite{zhd99b}. Its
general solution $\vec \omega = \vec \omega(t, \vec x)$ is given
implicitly within the equivalence relation (\ref{1.equ}) by the
following formulas:
\begin{equation}
\label{2.7} \vec x ={\cal O}(t){\cal L}(t)\left(\vec z(\vec
\omega) + \vec v(t)\right),
\end{equation}
Here ${\cal O}(t)$ is a time-dependent $3\times 3$ orthogonal
matrix with Euler angles $\alpha(t), \beta(t), \gamma(t)$:
\begin{equation} {\cal O}(t)= \left(\begin{array}{ccc} \cos \alpha\, \cos
\beta - \sin \alpha\, \sin \beta \cos \gamma & \quad -\cos
\alpha\, \sin \beta - \sin \alpha\, \cos \beta \cos \gamma & \quad
\sin \alpha\, \sin \gamma\\ \sin \alpha\, \cos \beta + \cos
\alpha\, \sin \beta \cos \gamma & -\sin \alpha\, \sin \beta + \cos
\alpha\, \cos \beta \cos \gamma & -\cos \alpha\, \sin \gamma\\
\sin \beta\, \sin \gamma & \cos \beta\, \sin \gamma & \cos \gamma
\end{array}\right);\label{2.m}
\end{equation}
$\vec v(t)$ stands for the vector-column whose entries $v_1(t),
v_2(t), v_3(t)$ are arbitrary smooth functions of $t$; $\vec z=
\vec z(\vec\omega)$ is given by one of the eleven formulas
\begin{eqnarray}
&1.\ & \mbox{Cartesian coordinate system,} \nonumber\\
&&z_1=\omega_1,\quad z_2=\omega_2,\quad z_3=\omega_3,\ \nonumber\\
&&\quad \omega_1,\omega_2,\omega_3\in {\bf R}.\ \nonumber\\ &2.\ &
\mbox{Cylindrical coordinate system,} \nonumber\\
&&z_1=e^{\omega_1} \cos \omega_2,\quad z_2=e^{\omega_1} \sin
\omega_2,\quad z_3=\omega_3,\ \nonumber\\ &&\quad
0\leq\omega_2<2\pi,\quad\omega_1,\omega_3\in {\bf R}.\ \nonumber\\
&3.\ & \mbox{Parabolic cylindrical coordinate system,} \nonumber\\
&&z_1=(\omega_1^2 - \omega_2^2)/2,\quad z_2=\omega_1
\omega_2,\quad z_3=\omega_3,\ \nonumber\\ &&\quad\omega_1>0,\quad
\omega_2,\omega_3\in{\bf R}.\ \nonumber\\ &4.\ & \mbox{Elliptic
cylindrical coordinate system,}\ \nonumber\\ &&z_1=a \cosh
\omega_1 \cos \omega_2,\quad z_2=a \sinh \omega_1 \sin
\omega_2,\quad z_3=\omega_3,\ \nonumber\\ &&\quad\omega_1>0,\quad
-\pi<\omega_2\leq \pi,\quad \omega_3\in{\bf R},\quad a>0.\
\nonumber\\ &5.\ & \mbox{Spherical coordinate system,} \nonumber\\
&&z_1=\omega_1^{-1}\, {\rm sech}\, \omega_2 \cos
\omega_3,\nonumber\\ &&z_2=\omega_1^{-1}\, {\rm sech}\, \omega_2
\sin \omega_3,\nonumber\\ &&z_3=\omega_1^{-1} \tanh \omega_2,\
\nonumber\\ &&\quad \omega_1>0,\quad \omega_2\in{\bf R},\quad
0\leq \omega_3<2\pi.\nonumber\\ &6.\ & \mbox{Prolate spheroidal
coordinate system,} \nonumber\\ &&z_1=a\, {\rm csch}\, \omega_1\,
{\rm sech}\, \omega_2 \cos \omega_3,\quad a>0,\nonumber\\
&&z_2=a\, {\rm csch}\, \omega_1\, {\rm sech}\, \omega_2 \sin
\omega_3,\nonumber\\ &&z_3=a \coth \omega_1 \tanh \omega_2,\
\label{1.2}\\ &&\quad\omega_1>0,\quad\omega_2\in{\bf R},\quad
0\leq \omega_3<2\pi.\nonumber\\ &7.\ & \mbox{Oblate spheroidal
coordinate system,} \nonumber\\ &&z_1=a \csc \omega_1\, {\rm
sech}\, \omega_2 \cos \omega_3,\quad a>0,\nonumber\\ &&z_2=a \csc
\omega_1\, {\rm sech}\, \omega_2 \sin \omega_3,\nonumber\\ &&z_3=a
\cot \omega_1 \tanh \omega_2,\ \nonumber\\ &&\quad
0<\omega_1<\pi/2,\quad\omega_2\in{\bf R},\quad
0\leq\omega_3<2\pi.\nonumber\\ &8.\ & \mbox{Parabolic coordinate
system,} \nonumber\\ &&z_1=e^{\omega_1 + \omega_2} \cos
\omega_3,\quad z_2=e^{\omega_1 + \omega_2} \sin
\omega_3,\nonumber\\ &&z_3=(e^{2 \omega_1} - e^{2
\omega_2})/2,\nonumber\\ &&\quad \omega_1,\omega_2\in {\bf
R},\quad 0\leq\omega_3\leq 2\pi.\nonumber\\ &9.\ & \mbox{
Paraboloidal coordinate system,} \nonumber\\ &&z_1=2 a \cosh
\omega_1 \cos \omega_2 \sinh \omega_3,\quad a>0,\nonumber\\ &&z_2=
2 a \sinh \omega_1 \sin \omega_2 \cosh \omega_3,\nonumber\\
&&z_3=a (\cosh 2\omega_1 + \cos 2\omega_2 - \cosh
2\omega_3)/2,\nonumber\\ &&\quad \omega_1,\omega_3\in{\bf R},\quad
0\leq\omega_2<\pi.\nonumber\\ &10.\ & \mbox{Ellipsoidal coordinate
system,} \nonumber\\ &&z_1=a\, {1\over {\rm sn} (\omega_1,\, k)}\,
{\rm dn} (\omega_2,\, k')\, {\rm sn} (\omega_3,\, k), \quad
a>0,\nonumber\\ &&z_2=a\, {{\rm dn} (\omega_1,\, k)\over {\rm sn}
(\omega_1,\, k)}\, {\rm cn} (\omega_2,\, k')\, {\rm cn}
(\omega_3,\, k), \nonumber\\ &&z_3=a\, {{\rm cn} (\omega_1,\,
k)\over {\rm sn} (\omega_1,\, k)}\, {\rm sn} (\omega_2,\, k')\,
{\rm dn} (\omega_3,\, k),\nonumber\\ &&\quad 0<\omega_1<K,\quad
-K'\leq \omega_2\leq K',\quad 0\leq \omega_3\leq 4K.\nonumber\\
&11.\ & \mbox{Conical coordinate system,} \nonumber\\
&&z_1=\omega_1^{-1} {\rm dn} (\omega_2,\, k')\, {\rm sn}
(\omega_3,\, k), \nonumber\\ &&z_2=\omega_1^{-1} {\rm cn}
(\omega_2,\, k')\, {\rm cn} (\omega_3,\, k),\nonumber\\
&&z_3=\omega_1^{-1} {\rm sn} (\omega_2,\, k')\, {\rm dn}
(\omega_3,\, k),\nonumber\\ &&\quad \omega_1>0,\quad -K'\leq
\omega_2\leq K',\quad 0\leq \omega_3\leq 4K;\nonumber
\end{eqnarray}
and ${\cal L}(t)$ is a $3\times 3$ diagonal matrix
\begin{equation}
\label{0}
{\cal L}(t) =\left(\begin{array}{ccc}
l_1(t) & 0 & 0\\0 & l_2(t) & 0\\0 & 0 & l_3(t)
\end{array}\right),
\end{equation}
where $l_1(t), l_2(t), l_3(t)$ are arbitrary non-zero smooth
functions that satisfy the following conditions
\begin{itemize}
\item $l_1(t) = l_2(t)$ for the partially split coordinate systems
    (cases 2--4 from (\ref{1.2})),
\item $l_1(t) = l_2(t) = l_3(t)$ for non-split coordinate systems (cases
    5--11 from (\ref{1.2})).
\end{itemize}
Here we use the usual notations for the trigonometric, hyperbolic
and Jacobi elliptic functions, number $k\, (0<k<1)$ being the
modulus of the latter and $k'=(1-k^2)^{1/2}$.

From a geometric point of view the right-hand side of formula
(\ref{2.7}) is a result of application to vector $\vec z(\vec
\omega)$ of the following time-dependent transformations performed
one after another:
\begin{enumerate}
\item
translations $\vec z\to \vec z'=\vec z+\vec v(t)$,
\item
dilatations $\vec z\to \vec z'={\cal L}(t)\vec z$,
\item
three-dimensional rotations $\vec z\to \vec z'={\cal O}(t)\vec z$
with Euler angles $\alpha(t)$, $\beta(t)$, $\gamma(t)$.
\end{enumerate}
Together with the rotations the following vector $\vec
\Omega(t)=(\Omega_1, \Omega_2, \Omega_3)$  is considered
\cite[\S35]{lanI}
\begin{eqnarray}
&&\Omega_1(t)=\dot\gamma(t) \cos\alpha(t) + \dot\beta(t)
\sin\alpha(t) \sin\gamma(t),\nonumber\\
&&\Omega_2(t)=\dot\gamma(t) \sin\alpha(t) - \dot\beta(t)
\cos\alpha(t) \sin\gamma(t),\label{kso}\\
&&\Omega_3(t)=\dot\alpha(t) + \dot\beta(t) \cos\gamma(t),\nonumber
\end{eqnarray}
that is directed along momentary axis of rotation and called {\it
angular velocity vector}.

Note that we have chosen the coordinate systems $\omega_1,
\omega_2, \omega_3$ by means of the equivalence relation
(\ref{1.equ}) in such a way that the relations
\begin{equation}
\label{1.3} \Delta \omega_a=0,\quad a=1,2,3
\end{equation}
hold for all the cases 1--11 in (\ref{1.2}).

After integration of system $(i)$--$(iii)$ it is not difficult to
integrate the remaining equations $(iv)$ and $(v)$ from the system
under study, since they can be regarded as algebraic equations for
the functions $A_a(t,\vec x), (a=1,2,3)$ and $A_0(t,\vec x)$,
correspondingly.

Multiplying equation $(iv)$ from the right by $Q^{-1}$ we obtain
for each component of matrices $\displaystyle{\partial Q \over
\partial x_b}Q^{-1}$, $b=1,2,3$ the systems of three linear
algebraic equations. The determinants of the systems do not vanish
according to (\ref{3a}). So, they have the unique solution
\begin{equation}
\label{p1} {\partial Q \over \partial x_b}Q^{-1}=f_b(t,\vec
x)I,\quad b=1,2,3,
\end{equation}
where $f_b(t,\vec x)$ are scalar smooth functions and $I$ is unit
$2\times 2$--matrix. From the compatibility conditions
\[
{\partial f_a\over \partial x_b}={\partial f_b\over \partial
x_a},\quad a,b=1,2,3
\]
of the above system of PDEs we obtain that there exists such
function $g(t,\vec x)$ that the equalities $f_a=\partial
g/\partial x_a$, $a=1,2,3$ hold. So (\ref{p1}) takes the form
\[
{\partial Q \over \partial x_b}={\partial g \over \partial
x_b}Q,\quad b=1,2,3.
\]
The general solution of this system of matrix PDEs is
\begin{equation}
\label{p2}
Q={\cal U}(t) \exp g(t,\vec x),
\end{equation}
where ${\cal U}(t)$ is arbitrary $2\times 2$--matrix function of
$t$.

Let us represent the complex-valued function $g(t,\vec x)$ in
(\ref{p2}) as $g=S_1+iS$, where $S_1, S$ are real-valued
functions. Now, if we take into account that the components of the
vector potential $A(t, \vec x)$ and functions $\omega_1,\omega_2,
\omega_3$ are real-valued functions, then after inserting
(\ref{p2}) into $(iv)$ with the use of (\ref{1.3}) we can split
the obtained equations into real and imaginary parts:
\begin{eqnarray}
&& {\partial S_1 \over \partial x_b}{\partial \omega_a\over
\partial x_b} = 0,\quad a=1,2,3; \label{3.2}\\ && 2\left({\partial
S \over
\partial x_b}- e A_b\right){\partial \omega_a\over \partial x_b} +
   {\partial\omega_a\over \partial t} = 0,\quad a=1,2,3.\label{3.1}
\end{eqnarray}
Taking into account the equality (\ref{3a}), we obtain from
(\ref{3.2}) the equalities $\partial S_1/\partial x_b=0,\
b=1,2,3$. It gives that $S_1=S_1(t)$.

Let us denote $e\vec{\cal A}=e\vec A-\vec\nabla S$. Then the
system (\ref{3.1}) takes the form of three linear algebraic
equations for functions ${\cal A}_1$, ${\cal A}_2$, ${\cal A}_3$:
\[
{\partial \omega_a \over \partial t}=2 e {\partial \omega_a \over
\partial x_b}{\cal A}_b,\quad a=1,2,3.
\]
The determinant of this system does not vanish due to (\ref{3a}).
Consequently, it has a unique solution. Making in this solution
the hodographic transformation
\begin{equation}
t=t,\quad x_a=u_a(t,\omega_1,\omega_2,\omega_3),\quad
a=1,2,3,\label{hod}
\end{equation}
we get the following expressions
for ${\cal A}_1$, ${\cal A}_2$, ${\cal A}_3$:
\[
\vec {\cal A} = -\frac{1}{2e}{\partial \vec u(t,\vec \omega) \over
\partial t}.
\]
After substitution into this formula expression for $\vec u(t,\vec
\omega)$ (\ref{2.7}), we return to variables $t,x_1,x_2,x_3$ and
thus obtain the following system:
\begin{equation}
\label{3.3a}
2(-e\vec A (t, \vec x)+\vec\nabla S)= {\cal M}(t)\vec x +
{\cal O}(t) {\cal L}(t)\dot{\vec v}.
\end{equation}
Here we use the designation
\begin{equation}
\label{m}
{\cal M}(t) =\dot{\cal O}(t) {\cal O}^{-1}(t) + {\cal O}(t)
\dot {\cal L}(t) {\cal L}^{-1}(t) {\cal O}^{-1}(t),
\end{equation}
where ${\cal O}(t),\ {\cal L}(t)$ are variable $3\times 3$ matrices defined
by formulas (\ref{2.m}) and (\ref{0}), correspondingly, $\vec v = (v_1(t),
v_2(t), v_3(t))^T$.
Note that $\dot{\cal O} {\cal O}^{-1}$ is antisymmetric
and  ${\cal O}\dot {\cal L} {\cal L}^{-1} {\cal O}^{-1}$
is symmetric part of matrix ${\cal M}$.

The direct calculation shows that equations $(v)$ and (\ref{3.3a})
are invariant under gauge transformations (\ref{calib2}). Thus the
function $S$ is transformed by the rule
\begin{equation}
S\to S'=S+ef,   \label{calib3'}
\end{equation}
which follows from (\ref{calib3}). In other words, if the
transformations (\ref{calib2}), (\ref{calib3'}) in equations $(v)$
and (\ref{3.3a}) are performed one after another, we obtain the
initial equations where functions $\vec A, A_0, S$ should be
replaced with functions $\vec A', A'_0, S'$. So, if the Pauli
equation (\ref{1.1}) with potential $\vec A, A_0$ admits
separation of variables in some coordinate system, then the Pauli
equation with potential $\vec A', A'_0$ admits separation of
variables in the same coordinate system (the multiplier $Q$
(\ref{p2}) is changed only). Therefore, it is worthwhile to fix
some gauge and to work only with representatives of the
equivalence classes of potentials $A(t, \vec x)$ (in the sense of
equivalence relation (\ref{calib2})).

We choose the gauge in a way that the equality
\begin{equation}
\label{s1}
2\vec\nabla S= {\cal O}(t) \dot {\cal L}(t) {\cal L}^{-1}(t)
{\cal O}^{-1}(t)\vec x + {\cal O}(t) {\cal L}(t)\dot{\vec v}.
\end{equation}
holds. After integration of this system of PDEs we obtain the
expression for $S$:
\begin{equation}
S={1\over 4}\sum\limits_{a=1}^{3}\left({\dot l_a\over l_a}{x'}_a^2
+2l_a\dot v_ax'_a\right), \label{S}
\end{equation}
where we use the notations
\begin{equation}
\label{x'}
\vec x'={\cal O}^{-1}\vec x.
\end{equation}

Next, we obtain from equation (\ref{3.3a}) the explicit form for
space-like components of vector-potential of electromagnetic field
\begin{equation}
\label{3.3.1}
\vec A (t, \vec x)= -\frac{1}{2e}\dot{\cal O} {\cal O}^{-1}\vec x,
\end{equation}
where the explicit form of matrix $\dot{\cal O} {\cal O}^{-1}$ is
given by the formula
\begin{eqnarray}
\dot{\cal O}{\cal O}^{-1} &=& \left(\begin{array}{ccc} 0 &
-(\dot\alpha + \dot\beta \cos\gamma) & \dot\gamma
\sin\alpha-\dot\beta \cos\alpha \sin\gamma\\
      \dot\alpha + \dot\beta \cos\gamma & 0 &-(\dot\gamma \cos\alpha+\dot\beta \sin\alpha \sin\gamma)\\
-(\dot\gamma \sin\alpha-\dot\beta \cos\alpha \sin\gamma) &
  \dot\gamma \cos\alpha+\dot\beta \sin\alpha \sin\gamma & 0
\end{array}\right), \label{anti}
\end{eqnarray}
where $\alpha, \beta, \gamma$ are arbitrary functions of $t$.

Thus formula (\ref{3.3.1}) means that the space-like components of
electromagnetic field $A(t, \vec x)$ are linear with respect of
spatial variables. So the magnetic field $\vec H={\rm rot}\ \vec
A$ should be homogeneous, i.e., independent of spatial variables
$\vec x$. From formulas (\ref{3.3.1}), (\ref{anti}) we can obtain
its explicit form
\begin{eqnarray}
&&eH_1=-\dot\gamma(t) \cos\alpha(t) - \dot\beta(t) \sin\alpha(t)
\sin\gamma(t),\nonumber\\ &&eH_2=-\dot\gamma(t) \sin\alpha(t) +
\dot\beta(t) \cos\alpha(t) \sin\gamma(t),\label{3.4}\\
&&eH_3=-\dot\alpha(t) - \dot\beta(t) \cos\gamma(t).\nonumber
\end{eqnarray}
Now the space-like components of the electromagnetic field take
the final form
\begin{equation}
\label{3.3}
\vec A (t, \vec x)=\frac{1}{2}
\left(\begin{array}{ccc}
0 & -H_3(t) & H_2(t)\\
H_3(t)& 0 &-H_1(t)\\
-H_2(t) & H_1(t)& 0
\end{array}\right)\vec x=\frac{1}{2}\, \vec H(t)\times \vec x,
\end{equation}
where symbol $\times$ denotes cross product.

Within the equivalence relation (\ref{0.6b}) we can always choose
the function ${\cal U}(t)$ to be a solution of matrix ODE
\begin{equation}
\label{U} i\dot{\cal U}=(-e\vec\sigma\vec H(t)){\cal U}
\end{equation}
with the initial conditions ${\cal U}(0)=I$. Due to the theorem of
existence and the uniqueness of the solution of the Cauchy problem
for the system of ODEs there is unique solution ${\cal U}(t)$ of
system (\ref{U}) for each fixed configuration of magnetic field
$\vec H(t)$. Moreover, matrix ${\cal U}(t)$ is a unitary one.
Indeed, taking into account (\ref{U}), we have the equality
\[
{d\over dt}({\cal U}^*{\cal U})= {\cal U}^*(ie\vec\sigma\vec
H){\cal U}+{\cal U}^*(-ie\vec\sigma\vec H){\cal U}=0,
\]
i.e., ${\cal U}^*{\cal U}=$ const. The initial conditions give\
${\cal U}^*{\cal U}=I$.

Thus we can consider the following change of variables in the
Pauli equation (\ref{1.1})
\begin{equation}
\label{pts}
\psi={\cal U}(t)\tilde\psi.
\end{equation}
Due to the unitarity of the matrix ${\cal U}$ the quantity
$\psi^*\psi$, which is regarded in quantum mechanics as the
probability density, is not changed. So the change of variables
(\ref{pts}) is the correct one. As a result, the term
$e\vec\sigma\vec H$ in Pauli equation (\ref{1.1}) vanishes, and we
obtained a system of two Schr\"odinger equations for the function
$\tilde\psi$.

Thus we proved the following assertion.
\begin{lemma}
A necessary condition for the Pauli equation (\ref{1.1}) to be
separable (in the sense of definition \ref{oz1}) is that it has to
be equivalent (in the sense of equivalence relation (\ref{pts}))
to a system of two uncoupled Schr\"odinger equations.
\end{lemma}

Let us substitute the equality (\ref{p2}) into equation $(v)$,
taking into account equations (\ref{U}) and $S_1=S_1(t)$.
Splitting the equation obtained into real and imaginary parts
(note that all functions $F_{00}, F_{a0}$, $a=1,2,3$ are
real-valued ones), we obtain the equalities
\begin{eqnarray}
  && \sum\limits_{a=1}^{3}\, F_{a0}(\omega_a) {\partial \omega_a\over \partial
    x_b}{\partial \omega_a\over \partial x_b} - {\partial S\over \partial t} -
    {\partial S\over \partial x_b}{\partial S\over \partial x_b}+2  e A_b {\partial S \over \partial x_b}-F_{00}(t) - e A_0 - e^2 A_bA_b=0, \label{A0}\\
  && \dot S_1 + \Delta S - e {\partial A_b\over \partial x_b}=0. \label{iA0}
 \end{eqnarray}
Inserting into equation (\ref{A0}) expressions for  $S$ (\ref{S})
and $A_1, A_2, A_3$ (\ref{3.3}), we obtain the explicit form of
$A_0$:
\begin{equation}
\label{3.5} eA_0(t,\vec x)=\sum\limits_{a=1}^{3}\,
F_{a0}(\omega_a) {\partial \omega_a \over \partial x_b}{\partial
\omega_a\over \partial x_b} -F_{00}(t) - e^2A_bA_b - {1\over 4}P.
\end{equation}
Here $A_bA_b$ follows from (\ref{3.3}), (\ref{anti}):
\begin{equation}
4A_bA_b=(H_2x_3-H_3x_2)^2+(H_3x_1-H_1x_3)^2+
(H_2x_1-H_1x_2)^2,\label{A^2}
\end{equation}
where $H_1,H_2,H_3$ are components of magnetic field (\ref{3.4});
function $P$ has the form
\begin{equation}
P=\sum\limits_{a=1}^{3}\left({\ddot l_a\over l_a}{x'}_a^2
+2(l_a\ddot v_a+2\dot l_a\dot v_a)x'_a+l_a^2\dot v_a^2\right),
\label{P}
\end{equation}
where $x'_1,x'_2,x'_3$ are given by formula (\ref{x'}) and
$l_a=l_a(t)$, $v_a=v_a(t)$, $a=1,2,3$, are arbitrary smooth
functions, which define new coordinate system (\ref{2.7}).

Let us emphasize that the expression for $A_0$ includes arbitrary
functions $F_{10}(\omega_1)$, $F_{20}(\omega_2)$,
$F_{30}(\omega_3), F_{00}(t)$, where functions
$\omega_a=\omega_a(t, \vec x)$, $a=1,2,3$, belong to one of 11
classes, whose representatives are given implicitly by the
formulas (\ref{2.7})--(\ref{0}).

Below we give explicit forms of the eikonals
$R_a^{-2}=\displaystyle{\partial \omega_a \over \partial
x_b}\displaystyle{\partial \omega_a\over \partial x_b}$ for each
class of $\omega_a$ (see, also \cite{zhd99b}):
\begin{eqnarray}
  &1.\ &R_i^{-2} =h_i^{-2},\quad i=1,2,3;\nonumber\\
  &2.\ &R_1^{-2} =R_2^{-2} =h_1^{-2} e^{-2\omega_1},\quad R_3^{-2} =h_3^{-2};\nonumber\\
  &3.\ &R_1^{-2} =R_2^{-2} =h_1^{-2} (\omega_1^2+\omega_2^2)^{-1},\quad R_3^{-2}
  =h_3^{-2};\nonumber\\
  &4.\ &R_1^{-2} =R_2^{-2} =h_1^{-2} a^{-2}(\cosh^2\omega_1 - \cos^2\omega_2)^{-1},
  \quad R_3^{-2} =h_3^{-2};\nonumber\\
  &5.\ &R_1^{-2} =h_1^{-2} \omega_1^4,\quad R_2^{-2} =R_3^{-2} =h_1^{-2} \omega_1^2
  \cosh^2\omega_2;\nonumber\\
  &6.\ &R_1^{-2} =h_1^{-2} a^{-2} \sinh^2 \omega_1 (\sinh^{-2} \omega_1 +
  \cosh^{-2}\omega_2)^{-1},\nonumber\\
  &&R_2^{-2} =h_1^{-2} a^{-2} \cosh^2 \omega_2 (\sinh^{-2} \omega_1 +
  \cosh^{-2}\omega_2)^{-1},\nonumber\\
  &&R_3^{-2} =h_1^{-2} a^{-2} \sinh^2 \omega_1 \cosh^2 \omega_2;\nonumber\\
  &7.\ &R_1^{-2} =h_1^{-2} a^{-2}\sin^2 \omega_1 (\sin^{-2} \omega_1 -
  \cosh^{-2}\omega_2)^{-1},\nonumber\\
  &&R_2^{-2} =h_1^{-2} a^{-2}\cosh^2 \omega_2 (\sin^{-2} \omega_1 -
  \cosh^{-2}\omega_2)^{-1},\label{R}\\
  &&R_3^{-2} =h_1^{-2} a^{-2}\sin^2 \omega_1 \cosh^2 \omega_2;\nonumber\\
  &8.\ &R_1^{-2} =h_1^{-2} e^{-2\omega_1} ( e^{2\omega_1} + e^{2\omega_2})^{-1},\nonumber\\
  &&R_2^{-2} =h_1^{-2} e^{-2\omega_2} ( e^{2\omega_1} + e^{2\omega_2})^{-1},\quad
  R_3^{-2}=h_1^{-2}e^{-2(\omega_1+\omega_2)};\nonumber\\
  &9.\ &R_1^{-2} =h_1^{-2} a^{-2}(\cosh 2\omega_1 - \cos 2\omega_2)^{-1}
  (\cosh 2\omega_1 + \cosh2\omega_3)^{-1},\nonumber\\
  &&R_2^{-2} =h_1^{-2} a^{-2}(\cosh 2\omega_1 - \cos 2\omega_2)^{-1}
  (\cos 2\omega_2 + \cosh2\omega_3)^{-1},\nonumber\\
  &&R_3^{-2} =h_1^{-2} a^{-2}(\cosh 2\omega_1 + \cosh 2\omega_3)^{-1}
  (\cos 2\omega_2 + \cosh2\omega_3)^{-1};\nonumber\\
  &10.\ &R_1^{-2} =h_1^{-2} a^{-2}\left ({{\rm dn}^2 (\omega_1,k)\over {\rm sn}^2 (\omega_1,k)}
    - k'^2 {\rm cn}^2 (\omega_2,k')\right)^{-1} \left({{\rm dn}^2 (\omega_1,k)\over {\rm sn}^2 (\omega_1,k)}
    + k^2 {\rm cn}^2 (\omega_3,k)\right)^{-1},\nonumber\\
  &&R_2^{-2} =h_1^{-2} a^{-2}\left({{\rm dn}^2 (\omega_1,k)\over {\rm sn}^2
    (\omega_1,k)} - k'^2 {\rm cn}^2 (\omega_2,k')\right)^{-1} \left(k'^2 {\rm cn}^2 (\omega_2,k') + k^2 {\rm cn}^2 (\omega_3,k)\right)^{-1},\nonumber\\
  &&R_3^{-2} =h_1^{-2} a^{-2}\left({{\rm dn}^2 (\omega_1,k)\over {\rm sn}^2
    (\omega_1,k)} + k^2 {\rm cn}^2 (\omega_3,k)\right)^{-1} \left(k'^2 {\rm cn}^2 (\omega_2,k') + k^2 {\rm cn}^2 (\omega_3,k)\right)^{-1};\nonumber\\
  &11.\ &R_1^{-2} =h_1^{-2} \omega_1^4,\quad R_2^{-2} =R_3^{-2} =h_1^{-2}\omega_1^2 \left(k'^2 {\rm cn}^2 (\omega_2,k') +
  k^2 {\rm cn}^2 (\omega_3,k)\right)^{-1}.\nonumber
\end{eqnarray}

At last, let us find the multiplier $Q$. Substituting the formulas
(\ref{S}) and (\ref{3.3}) into equation (\ref{iA0}) gives
\[
\dot S_1 = -{1 \over 2}\sum\limits_{a=1}^{3}\,{\dot l_a \over l_a},
\]
whence it follows that
\begin{equation}
\label{S1}
S_1 = -{1 \over 2}\sum\limits_{a=1}^{3}\,\ln l_a.
\end{equation}
Taking into account expression for $S$ (\ref{S}), we obtain from
formula (\ref{p2}) the explicit form of $Q$
\begin{equation}
\label{pQ}
Q={\cal U}(t){1\over \sqrt{l_1l_2l_3}}\exp\sum\limits_{a=1}^{3}{i\over 4}
\left({\dot l_a\over l_a}{x'}_a^2 + 2l_a\dot v_ax'_a\right),
\end{equation}
where ${\cal U}(t)$ is given by the equation (\ref{U}), and
$x'_1,x'_2,x'_3$ are given by formula (\ref{x'}).

Thus we have proved the main result of the article:
\begin{theorem}
\label{the3} Pauli equation (\ref{1.1}) admits separation of
variables (in the sense of definition \ref{oz1}) if and only if it
is gauge equivalent to Pauli equation where
\begin{enumerate}
\item[--]
the magnetic field $\vec H={\rm rot}\ \vec A$ is independent of
the spatial variables,
\item[--]
the space--like components $A_1$, $A_2$, $A_3$ of the
vector--potential of the electromagnetic field are given by
(\ref{3.3}),
\item[--]
the time--like component $A_0$ is given by formulas
(\ref{3.5})--(\ref{R}).
\end{enumerate}
\end{theorem}

Comparing the components of magnetic field (\ref{3.4}) with
components of angular velocity vector (\ref{kso}) of rotation of
coordinate system (\ref{2.7}), we obtain the equality $e\vec H =
-\vec \Omega$. So, we prove the following assertion:
\begin{corollary}
Let Pauli equation (\ref{1.1}) admit separation of variables in
some non-stationary coordinate system $t$, $\omega_a=\omega_a(t,
\vec x)$, $a=1,2,3$, where functions $\omega_1(t, \vec x)$,
$\omega_2(t, \vec x)$, $\omega_3(t, \vec x)$ are given implicitly
by formulas (\ref{2.7})--(\ref{0}). Then angular velocity vector
(\ref{kso}) of rotation of this coordinate system equals $-e\vec
H$, where $\vec H={\rm rot}\ \vec A$ is magnetic field.
\end{corollary}

It follows from the corollary that a necessary condition for the
Pauli equation (\ref{1.1}) with non-zero magnetic field $\vec H$
to be separable (in the sense of our definition 1) is that the
angular velocity vector (\ref{kso}) of rotation of the separation
coordinate system (\ref{2.7})--(\ref{0}) has to be non-zero.

Summing up we conclude that coordinate systems and
vector-potentials of the electromagnetic field $A(t, \vec
x)=(A_0(t, \vec x), \vec A(t, \vec x))$ providing separability of
the corresponding Pauli equations coincide with those for the
Schr\"odinger equations. Namely, we prove that the magnetic field
$\vec H={\rm rot}\ \vec A$ has to be independent of the spatial
variables. Next, we have eleven classes of potentials $A_0(t, \vec
x)$, corresponding to eleven classes of coordinate systems
$t,\omega_a=\omega_a(t, \vec x),\ a=1,2,3$, where the functions
$\omega_1(t, \vec x), \omega_2(t, \vec x), \omega_3(t, \vec x)$
are given implicitly by formulas (\ref{2.7})--(\ref{0}). Pauli
equation (\ref{1.1}) for each class of the functions $A_0(t, \vec
x), \vec A(t, \vec x)$ defined by (\ref{3.3}), (\ref{3.5}) and
(\ref{R}) under arbitrary $F_{00}(t), F_{a0}(\omega_a)$  and fixed
arbitrary functions $\alpha(t)$, $\beta(t)$, $\gamma(t)$,
$v_a(t)$, $l_a(t)$, $a=1,2,3$, separates in exactly one coordinate
system.

The solutions with separated variables are of the form
(\ref{2.1}), where $Q$ is given by (\ref{pQ}). The separation
equations read as (\ref{2.7a}) or (\ref{2.8a}), where the
functions $F_{\mu0}, \mu=0,1,2,3$, are arbitrary smooth functions
defining the form of the time-like component of the
vector-potential $A(t, \vec x)$ (see, (\ref{3.5})). The explicit
forms of other coefficients $F_{\mu a}, G_{\mu\nu}, G_\mu$ of
reduced equations can be obtained by splitting relations ($ii$)
and ($iii$) with respect to independent variables $\omega_1,
\omega_2, \omega_3, t$ for each class of the functions $\vec z =
\vec z (\vec \omega)$ given in (\ref{1.2}). Let us denote
\begin{equation}
\label{stac}
S=\left(\begin{array}{ccc} T_1 & T_2 & T_3\\ S_{11} & S_{12} &
S_{13}\\ S_{21} & S_{22} & S_{23}\\ S_{31} & S_{32} & S_{33}
\end{array}\right),
\end{equation}
where the functions $S_{ab}(\omega_a)\ (a,b=1,2,3)$ are given
below as entries of $3\times 3$ St\"ackel matrices, whose
structure is determined by the choice of the functions $\vec z =
\vec z (\vec \omega)$:
\begin{eqnarray}
&&{\cal F}_1=\left(\begin{array}{ccc} 1 & 0 & 0 \\ 0 & 1 & 0\\ 0 &
0 & 1 \end{array}\right),\quad {\cal F}_2=
\left(\begin{array}{ccc} e^{2\omega_1} & -1 & 0\\ 0 & 1 & 0\\ 0 &
0 & 1 \end{array}\right),\quad {\cal F}_3=
\left(\begin{array}{ccc} \omega_1^2 & -1 & 0\\ \omega_2^2 & 1 &
0\\ 0 & 0 & 1
\end{array}\right),\nonumber\\[3mm]
&& {\cal F}_4= \left(\begin{array}{ccc} a^2 \cosh^2 \omega_1 & 1 &
0\\ -a^2 \cos^2\omega_2 & -1 & 0\\ 0 & 0 & 1
\end{array}\right),\quad {\cal F}_5= \left(\begin{array}{ccc}
\omega_1^{-4} & -\omega_1^{-2} & 0 \\ 0 & \cosh^{-2} \omega_2 &
-1\\ 0 & 0 & 1
\end{array}\right),\nonumber\\[3mm]
&&{\cal F}_6= \left(\begin{array}{ccc} a^2 \sinh^{-4} \omega_1 &
-\sinh^{-2} \omega_1 & -1 \\ a^2 \cosh^{-4} \omega_2 & \cosh^{-2}
\omega_2 & -1\\ 0 & 0 & 1 \end{array}\right),\quad {\cal
F}_7=\left(\begin{array}{ccc} a^2 \sin^{-4} \omega_1 & -\sin^{-2}
\omega_1 & 1 \\ -a^2 \cosh^{-4} \omega_2 & \cosh^{-2} \omega_2 &
-1\\ 0 & 0 & 1
\end{array}\right),\label{2.s}\\[3mm]
&& {\cal F}_8= \left(\begin{array}{ccc} e^{4\omega_1} &
-e^{2\omega_1} & -1 \\ e^{4\omega_2} & e^{2\omega_2} & -1\\ 0 & 0
& 1
\end{array}\right),\quad {\cal F}_9=
\left(\begin{array}{ccc} a^2 \cosh^2 2\omega_1 & -a\cosh 2\omega_1
& -1
\\ -a^2 \cos^2 2\omega_2 & a\cos 2\omega_2 & 1\\ a^2 \cosh^2 2\omega_3 &
a\cosh 2\omega_3 & -1
\end{array}\right),\nonumber\\[3mm]
&&{\cal F}_{10}= \left(\begin{array}{ccc} a^2 \displaystyle{{\rm
dn}^4 (\omega_1,k) \over {\rm sn}^4 (\omega_1,k)}&
-\displaystyle{{\rm dn}^2 (\omega_1,k) \over {\rm sn}^2
(\omega_1,k)}& 1
\\ -a^2 k'^4\, {\rm cn}^4 (\omega_2,k') & k'^2\, {\rm cn}^2
(\omega_2,k') & -1  \\ a^2 k^4\, {\rm cn}^4 (\omega_3,k) & k^2\,
{\rm cn}^2 (\omega_3,k)  & 1
\end{array}\right),\quad {\cal F}_{11}=\left(\begin{array}{ccc} \omega_1^{-4} & -\omega_1^{-2} & 0 \\
0 & k'^2 {\rm cn}^2 (\omega_2,k') & -1 \\ 0 & k^2 {\rm cn}^2
(\omega_3,k) & 1
\end{array}\right).\nonumber
\end{eqnarray}
The functions $T_1(t), T_2(t), T_3(t)$ are expressed in terms
of the functions $h_1(t)$, $h_2(t)$, $h_3(t)$:
\begin{eqnarray}
&1.\ &T_i =h_i^{-2},\quad i=1,2,3;\nonumber\\ &2-4.\ &T_1
=h_1^{-2},\quad T_2 =0,\quad T_3 =h_3^{-2};\label{5a}\\ &5-11.\
&T_1 =h_1^{-2},\quad T_2 =T_3 =0\nonumber.
\end{eqnarray}

Let $K$ and $M$ be $3\times 3$ constant matrices. Now, if the
reduced equations are given by (\ref{2.7a}), then
\[
F=\left \|F_{\mu a}\right\|_{\mu =0\; a=1}^{3\quad\; 3},\quad G=\left
\|G_{\mu a}\right\|_{\mu =0\; a=1}^{3\quad\; 3}
\]
are block $(6\times 8)$-matrices, where $F_{\mu a}$ and $G_{\mu
a}$ are $2\times 2$-matrices that are equal to products of the
corresponding entries of the matrices $SK$ and $SM$ by the unit
(in the case of the matrix $F$) or $\vec s\vec\sigma$ (in the case
of the matrix $G$) matrices. Accordingly, equation (\ref{2.3})
takes the form
\begin{equation}
\label{cond}
{\rm rank}\, (F+G)=6.
\end{equation}
If rank $K=3$, then we can always rearrange $\lambda_1, \lambda_2,
\lambda_3$ with the use of the equivalence relation (\ref{la.1})
in order to get $K=I$. Analogously, without loss of generality we
may put $M=I$, provided rank\,$M=3$ and $\vec s^2\ne 0$.

If rank $M=0$, then the column $\left \|G_{\mu 0}\right\|_{\mu =0}^{3}$
has necessarily the form $S\vec g$, where $\vec g$ is a constant
three-component column. If rank $M\ne 0$, then we can always kill this
column by a proper rearranging of $\lambda_1, \lambda_2, \lambda_3$
with the use of the equivalence relation (\ref{la.1}).

Next, if the reduced equations are given by (\ref{2.8a}), then the
matrix $F$ is defined in the same way as in the previous
case. Furthermore,
\[
G=\left \|G_\mu\vec s_a\vec\sigma\right\|_{\mu =0\; a=1}^{3\quad\;
3}
\]
is a block $(6\times 8)$-matrix, where $G_\mu$ are the
three-component columns $S\vec g_\mu$ ($\vec g_\mu$ is a constant
three-component column). In addition, in this case identity
(\ref{cond}) holds, so that we can put $K=I$, when rank $K=3$. If
$\vec s_a, (1=1,2,3)$ are three linear independent vectors, then
we can always put $\vec s_0=0.$

We will finish this section with the following remark. It follows
from theorem 1 that a choice of magnetic fields $\vec H$ allowing
for variable separation in the corresponding Pauli equation is
very restricted. Namely, the magnetic field should be independent
of spatial variables $x_1, x_2, x_3$ in order to provide the
separability of Pauli equation (\ref{1.1}) into three second-order
matrix ordinary differential equations of the form (\ref{2.2}).
However, if we allow for separation equations to be of a lower
order, then additional possibilities for variable separation in
the Pauli equation arise. As an example, we give the vector
potential
\[
A(t, \vec x)=\left(A_0\left(\sqrt{x_1^2+x_2^2}\right),\ 0,\ 0,\
A_3\left(\sqrt{x_1^2+x_2^2}\right)\right),
\]
where $A_0, A_3$ are arbitrary smooth functions. The Pauli equation
(\ref{1.1}) with this vector-po\-ten\-ti\-al separates in the
cylindrical coordinate system
\[
t,\quad \omega_1=\ln \left(\sqrt{x_1^2+x_2^2}\right),\quad
\omega_2=\arctan(x_1/x_2),\quad \omega_3=x_3
\]
into two first-order and one second-order matrix ordinary
differential equations. The corresponding magnetic field $\vec
H={\rm rot}\, \vec A$ is evidently $x$-dependent. In this respect,
let us also mention the recent paper by Benenti with co-authors
\cite{ben01}, where the problem of separation of variables in the
stationary Hamilton-Jacobi equation with vector-potential has been
studied. They have presented a number of vector-potentials, for
which the Hamilton-Jacobi equation is separable, and the
corresponding magnetic fields are inhomogeneous ones. These
potentials allow for separation of variables in the stationary
Schr\"odinger and Pauli equations with vector-potentials as well
(see, e.g., \cite{koo80} concerning the relationship between the
separation of variables in the Schr\"odinger and Hamilton-Jacobi
equations). These facts imply an importance of application of our
approach to classify the non-stationary Pauli equations of the
form (\ref{1.1}), which admit separation of variables into first-
and second-order matrix ordinary differential equations. We remind
that here we give the classification results for the case, when
all the reduced equations are second-order ones. We intend to
address this problem in one of our future publications.

\section{Algorithm of separation of variables in the Pauli
equation with fixed potential}

Theorem \ref{the3} gives the solution of the problem of
classification of the Pauli equations (\ref{1.1}) with variable
coefficients that are separable (in the sense of definition 1) at
least in one coordinate system.

Let us consider the problem of classification of coordinate
systems that allow for separation of variables (in the sense of
definition 1) in the Pauli equation (\ref{1.1}) with fixed
vector-potential $A_0, \vec A$.

Let some fixed vector-potential $\vec A(t, \vec x), A_0(t, \vec
x)$ be given. The scheme of finding all coordinate systems
providing separation of variables is as follows:
\begin{enumerate}
\item{
With help of gauge transformations (\ref{calib2}) we reduce the
space-like components of vector-potential $\vec A (t, \vec x)$ to
the form (\ref{3.3}). If it is impossible, then Pauli equation
(\ref{1.1}) with this vector-potential is not solvable by the
method of separation of variables in the framework of our
approach.}

\item{
We solve the system of ODE (\ref{3.4}) for given magnetic field
$\vec H(t)$ and obtain the explicit form of functions $\alpha(t)$,
$\beta(t)$, $\gamma(t)$.}

\item{For each of 11 classes of coordinate systems
$t$, $\omega_a=\omega_a(t, \vec x)$, $a=1,2,3$, which are given by
formulas (\ref{2.7})--(\ref{0}), taking into account restrictions
obtained on the first step of the algorithm, we find the explicit
form of
\begin{enumerate}
\item
the time-like component $A_0$ of the vector-potential in terms of
$\vec \omega$;
\item
function $P$, substituting in (\ref{P}) the expression for $\vec
x'$ in terms of $\vec \omega$ (see formulas (\ref{x'}) and
(\ref{2.7})):
\begin{equation}
\label{x'1} \vec x' ={\cal L}(t)\left(\vec z(\vec \omega) + \vec
v(t)\right);
\end{equation}
\item
quantity $e^2A_bA_b$ by the formula
\[
4e^2A_bA_b=(n_2x'_3-n_3x'_2)^2+(n_3x'_1-n_1x'_3)^2+(n_2x'_1-n_1x'_2)^2,
\]
where $x'_1,x'_2,x'_3$ are given by the formula (\ref{x'1}), and
functions $n_1,n_2,n_3$ are as follows
\begin{eqnarray}
&&n_1=\dot\gamma(t) \cos\beta(t) + \dot\alpha(t) \sin\beta(t)
\sin\gamma(t),\nonumber\\ &&n_2=-\dot\gamma(t) \sin\beta(t) +
\dot\alpha(t) \cos\beta(t) \sin\gamma(t),\label{h}\\
&&n_3=\dot\beta(t) + \dot\alpha(t) \cos\gamma(t);\nonumber
\end{eqnarray}
\item
eikonals $\displaystyle{{\partial \omega_a \over \partial
x_b}{\partial \omega_a\over \partial x_b}}=R_a^{-2}$, $a=1,2,3$,
which are determined from the list (\ref{R}) for given class of
coordinates.
\end{enumerate}
}
\item{
We substitute the equalities obtained into equation (\ref{3.5})
and obtain 11 equations for each of 11 classes of coordinate
systems $t$, $\omega_1$, $\omega_2$, $\omega_3$. For each of these
equalities we find all possible functions $F_{a0}(\omega_a),\
a=1,2,3$, $F_{00}(t)$ that reduce it to the identity by the
independent variables $t$, $\omega_1$, $\omega_2$, $\omega_3$
(i.e. we split this equality with respect to these variables). It
gives, in its turn, the explicit form of the functions $v_a(t)$,
$l_a(t)$, $\ a=1,2,3$ and additional restriction on $\alpha(t)$,
$\beta(t)$, $\gamma(t)$, giving the form of the coordinate system
in question. All obtained coordinates for which the functions
$F_{a0}(\omega_a),\ a=1,2,3$, $F_{00}(t)$ exist are only
coordinate systems providing separability of Pauli equations in
the sense of definition \ref{oz1}.}
\end{enumerate}
{\bf Example.}\quad As illustration of this algorithm consider the
problem of separation of variables in Pauli equation (\ref{1.1})
for a particle interacting with a constant magnetic field. Without
loss of generality we can always choose it as directed along axes
$OZ$: $e\vec H=(0,0,c)^T$, where $c$ is a non-zero real constant.
The vector-potential of electro-magnetic field has the form
\begin{equation}
\label{10.1}
2e\vec A=\left(\begin{array}{ccc}
0 & -c & 0\\
c & 0 & 0\\
0 & 0 & 0 \end{array}\right) \vec x,\quad
eA_0={q\over |\vec x|} - {c^2\over 12}\left(x_1^2+x_2^2-2x_3^2\right),
\end{equation}
where $q$ is a non-zero real constant and $|\vec
x|=\sqrt{x_1^2+x_2^2+x_3^2}$.

A direct check shows this vector-potential satisfies the vacuum
Maxwell equations without currents \begin{eqnarray} &&\Box A_0
-\frac{\partial}{\partial t}\, \left(\frac{\partial A_0}{\partial
t} + {\rm div}\, \vec A\right)=0, \label{max}\\ &&\Box \vec A +
\vec{\rm grad}\, \left(\frac{\partial A_0}{\partial t}+ {\rm
div}\, \vec A\right)=\vec 0, \nonumber
\end{eqnarray}
where $\Box=\partial^2/\partial t^2 - \Delta$ is d'Alembert
operator. Therefore, it is a natural generalization of the
standard Coulomb potential, which is obtained from (\ref{10.1})
under $c\to 0$.
\begin{proposition}
The set of inequivalent coordinate systems providing separability
of the Pauli equation (\ref{1.1}) with vector potential of
electromagnetic field (\ref{10.1}) is exhausted by the following
ones:
\begin{equation}
\label{sk}
\vec x={\cal O}(t)\vec z,
\end{equation}
where ${\cal O}$ is a time-dependent $3\times 3$ orthogonal matrix
(\ref{2.m}), with Euler angles
\begin{equation}
\alpha(t)=-ct,\quad \beta={\rm const},\quad \gamma={\rm const},
\label{10.3}
\end{equation}
and $\vec z$ is one of the following coordinate systems:
\begin{enumerate}
\item
spherical (formula 5 from list (\ref{1.2})),
\item
prolate spheroidal II (formula 6 from (\ref{1.2}), where one
should replace $z_3$ with $z_3=a (\coth \omega_1 \tanh \omega_2\pm
1)$),
\item
conical (formula 11 from (\ref{1.2})).
\end{enumerate}
\end{proposition}
{\bf Proof.}$\quad$ The space-like component $\vec A (t, \vec x)$
of the given vector-potential (\ref{10.1}) is already reduced to
form (\ref{3.3}).

The system of ODE (\ref{3.4}) for given magnetic field takes the form:
\[
\dot\gamma \cos\alpha + \dot\beta \sin\alpha \sin\gamma=0,\quad
\dot\gamma \sin\alpha - \dot\beta \cos\alpha \sin\gamma=0,\quad
\dot\alpha + \dot\beta \cos\gamma =-c.
\]
This implies the equivalent system
\[
\dot\gamma=0,\quad\dot\beta\sin\gamma=0,
\quad\dot\alpha + \dot\beta \cos\gamma =-c.
\]
Its general solution up to translation by $t$ is given by formulas
(\ref{10.3}) (solution  $\alpha\pm\beta=-ct$ for case
$\sin\gamma=0$ is included into (\ref{10.3}) as a particular case
after denoting $\alpha\pm\beta\to \alpha$).

The steps 3 and 4 of the above algorithm will be illustrated by
the case of spherical coordinate system 5 from (\ref{1.2}) (for
other coordinate systems this procedure is an analogous one). For
this case the equality (\ref{3.5}) in terms
$\omega_1,\omega_2,\omega_3$ takes the form
\begin{eqnarray}
&&
l^{-2}\left(F_{10}(\omega_1)\omega_1^4+(F_{20}(\omega_2)+F_{30}(\omega_3))\omega_1^2\cosh^2\omega_2
\right) - F_{00}(t) = \nonumber \\ && ={q\over |\vec x'|} +
\left({c^2\over 6}+{1\over 4}{\ddot l\over l}\right) |\vec x'|^2 +
\sum\limits_{a=1}^{3}\left( 2(l\ddot v_a+2\dot l\dot
v_a)x'_a+l^2\dot v_a^2\right),\label{10.4}
\end{eqnarray}
where $l=l_1=l_2=l_3$, $l\neq 0$ (because of the spherical
coordinate system is non-split one), and
\begin{eqnarray}
&&x'_1=l(\omega_1^{-1}\, {\rm sech}\, \omega_2 \cos
\omega_3+v_1(t)),\nonumber\\ &&x'_2=l(\omega_1^{-1}\, {\rm sech}\,
\omega_2 \sin \omega_3+v_2(t)),\nonumber\\ &&x'_3=l(\omega_1^{-1}
\tanh \omega_2+v_3(t)).\ \nonumber
\end{eqnarray}
Next we perform on both parts of equality (\ref{10.4}) the
following step by step operations:
\begin{enumerate}
\item
multiplying by $\omega_1$,
\item
differentiation with respect to $\omega_1$,
\item
division by $\omega_1^2$,
\item
differentiation with respect to $\omega_1$,
\item
differentiation with respect to $\omega_2$,
\item
multiplying by $l\omega_1^7|\vec x'|^7$,
\item
twice multiplying by $\omega_1$.
\end{enumerate}
As result we get
\begin{eqnarray*}
&&-24q\, {\rm sech}^3\omega_2 (v_1^2+v_2^2+v_3^2) (v_1\cos
\omega_3+v_2\sin \omega_3+v_3\sinh\omega_2)\times\\
&&(-v_3+(v_1\cos\omega_3+v_2\sin\omega_3)\sinh\omega_2)=0.
\end{eqnarray*}
The equality obtained is transformed into an identity with respect
to independent variables $\omega_1$, $\omega_2$, $\omega_3$ if and
only if the condition $v_1=v_2=v_3=0$ holds. Now the equality
(\ref{10.4}) takes the form
\begin{equation}
\label{10.5}
l^{-2}\left(F_{10}(\omega_1)\omega_1^4+(F_{20}(\omega_2)+F_{30}(\omega_3))\omega_1^2\cosh^2\omega_2
\right) -F_{00}(t)= {q\over l}\omega_1 + \left({c^2\over
6}+{1\over 4}{\ddot l\over l}\right) {l^2\over \omega_1^2}.
\end{equation}

Performing on both parts of equality (\ref{10.5}) the following
step by step operations:
\begin{enumerate}
\item
differentiation with respect to $\omega_1$,
\item
multiplying by $l^2$,
\item
differentiation with respect to $t$,
\item
multiplying by $\omega_1^3$,
\item
differentiation with respect to $\omega_1$,
\end{enumerate}
we get the equality $3q\dot l\omega_1^2=0$. This implies $l=$
const and with the help of dilatations we can put without loss of
generality $l=1$. Thus the coordinate system takes the form
(\ref{sk}).

The equation (\ref{10.5}) yields
\[ F_{10}(\omega_1)\omega_1^4+(F_{20}(\omega_2)+F_{30}(\omega_3))\omega_1^2\cosh^2\omega_2
-F_{00}(t) =q\omega_1 + {c^2\over 6}\omega_1^{-2}.
\]
We can split the equation obtained by the independent variables
$\omega_1$, $\omega_2$, $\omega_3$. As a result we get
\begin{eqnarray*}
&& F_{10}=q\omega_1^{-3}+{c^2\over
6}\omega_1^{-6}+k_1\omega_1^{-4}- k_2\omega_1^{-2},\\ &&
F_{20}=k_2\, {\rm sech}^2\omega_2-k_3,\quad F_{30}=k_3,\quad
F_{00}=k_1.
\end{eqnarray*}
The theorem is proved. $\rhd$
\vspace{2mm}

\section{Separation of variables in the Pauli-Maxwell system}

The expressions (\ref{3.3}), (\ref{3.5})--(\ref{R}) give the most
general form of the vector-potential of the electromagnetic field,
providing separability of the corresponding Pauli equations. But,
because of generality of the results, these expressions are too
cumbersome, and their physical interpretation is somewhat
difficult. Therefore it would be interesting to know the form of
these potentials under certain physical restrictions. The most
natural restriction is that the vector-potential satisfies the
vacuum Maxwell equations without currents (\ref{max}).

In this section we describe all explicit forms of the
vector-potentials $A(t, \vec x)$ that
\begin{enumerate}
\item[a)] provide separability of Pauli equation,
\item[b)] satisfy vacuum Maxwell equations without currents
(\ref{max}) and
\item[c)] describe the non-zero magnetic field.
\end{enumerate}
Furthermore, we construct inequivalent coordinate systems enabling
us to separate variables in the corresponding Pauli equation.

The similar problem with more strong restrictions was analyzed in
\cite{zhd98a} for a two-dimen\-sional Schr\"odinger equation with
vector-potential. Note that an analogous problem for the Dirac
equation for an electron was analyzed in \cite{cook82}.

Taking into account the form of $\vec A$ (\ref{3.3}), the Maxwell
equations (\ref{max}) take the form
\begin{equation}
\Delta A_0=0,\label{10.10}
\end{equation}
and \[ {\partial^2 A_0\over \partial t\partial x_1}=-\ddot
l_3x_2+\ddot l_2x_3,\quad {\partial^2 A_0\over \partial t\partial
x_2}=\ddot l_3x_1-\ddot l_1x_1, \quad {\partial^2 A_0\over
\partial t\partial x_3}=-\ddot l_2x_1+\ddot l_1x_2.
\]
From the compatibility conditions of the above system of PDEs we
get
\begin{eqnarray}
&& \ddot l_1=\ddot l_2=\ddot l_2=0,\nonumber\\ && {\partial^2
A_0\over
\partial t\partial x_a}=0,\quad a=1,2,3. \label{10.2}
\end{eqnarray}

Inserting expression for potential $A_0(t, \vec x)$ (\ref{3.5})
into (\ref{10.10}) with subsequent change of independent variables
(\ref{2.7}) yields (we use the relations $\Delta \omega_i=0$,
$\omega_{ix_a}\omega_{jx_a}=0$, $i\not = j$, $i,j=1,2,3$)
\begin{equation}
\sum\limits_{j=1}^{3}\, \frac{\partial^2}{\partial
\omega_j^2}\left(\sum\limits_{i=1}^{3}\,
F_{i0}(\omega_i)R_i^{-2}\right)R_j^{-2}= {1\over
2}\sum\limits_{i=1}^{3}{\ddot l_i\over l_i}+
e^2(H_1^2+H_2^2+H_3^2), \label{5.1}
\end{equation}
where the eikonals
\begin{equation}
\label{5.2} R_i^{-2}={\partial \omega_i \over \partial
x_a}{\partial \omega_i\over \partial x_a}, \quad i=1,2,3
\end{equation}
are given in the list (\ref{R}).

Thus we get eleven functional relations ${\cal P}_1, \ldots, {\cal
P}_{11}$ for each class of coordinate system (\ref{2.7}), whose
form is determined by the form of one of the eleven expressions
$z_1(\omega_1, \omega_2, \omega_3)$, $z_2(\omega_1, \omega_2,
\omega_3)$, $z_3(\omega_1, \omega_2, \omega_3)$ from the list
(\ref{1.2}). As $t, \omega_1, \omega_2, \omega_3$ are functionally
independent, we can split the above relations with respect to the
variables $t, \omega_1, \omega_2, \omega_3$, thus getting ordinary
differential equations for the functions $F_{i0}(\omega_i),l_i(t),
i=1,2,3$. After solving them the formula (\ref{3.5}) yields the
expressions for $A_0$ in terms of variables $t, \omega_1,
\omega_2, \omega_3$. Returning to variables $t, x_1, x_2, x_3$
(with the aid of (\ref{2.7})), we should split the expression
obtained for $A_0(t, \vec x)$ with respect to $t$. Indeed, the
general solution of the equation (\ref{10.2}) is
\[
A_0(t, \vec x)=f_1(\vec x)+f_2(t).
\]
At the expense of the gauge invariance of the Pauli equation we
may choose $f_2(t)=0$. Thus the potential $A_0$ should be a
function of $\vec x$ only. This condition restricts the choice of
$A_0$, thus giving ordinary differential equations for the
functions $l_i(t),v_i(t), i=1,2,3$. Solving them we obtain the
explicit forms of the function $F_{00}(t)$ and coordinate systems
(\ref{2.7}). After simplifying these coordinate systems with the
aid of equivalence transformations we get a full description of
the vector-potentials $A(t, \vec x)$ and coordinate systems,
giving the solution of the problem under study.

Omitting the details of the calculations (they are very
cumbersome) we present below the results. Note, when presenting
lists of the vector-potentials $A(t, \vec x)$ and coordinate
systems we use invariance of the system of the Pauli and Maxwell
equations with respect to the groups of rotations by spatial
variables $x_1, x_2, x_3$ and translations by all variables $t,
x_1, x_2, x_3$ (see, e.g., \cite{nik94}). \vspace{4mm}

\noindent{\bf 1. Case of non-stationary magnetic field:}
\begin{eqnarray*}
&&e\vec H=\left(0,0,At+B\right),\\
&&eA_0=-{k\over 2}(x_1^2+x_2^2-2x_3^2) +a_1x_1+a_2x_2+a_3x_3,
\end{eqnarray*}
where $A,B,k,a_1,a_2,a_3$ are arbitrary real constants.

The coordinate system is
\[
\vec x= {\cal L O}(\vec z+\vec v).
\]
Here ${\cal O}$ is a time-dependent $3\times 3$ orthogonal matrix
${\cal O}(\alpha,\beta,\gamma)$, where
\[\alpha=-\displaystyle{1\over 2}At^2-Bt,\quad \beta=0,\quad \gamma=0  ;\]
$\vec z$ is Cartesian, cylindrical or elliptic cylindrical
coordinate system (formulas 1, 2, 4 from (\ref{1.2}); ${\cal L}$
is the $3\times 3$ diagonal matrix
\[
{\cal L}=\left(\begin{array}{ccc}
l(t) & 0 & 0\\0 & l(t) & 0\\0 & 0 & l_3(t)
\end{array}\right),
\]
and $\vec v(t)$ is vector-column $\vec v(t)=(v_1,v_2,v_3)^T$ where
functions $l(t),l_3(t),v_1(t),v_2(t), v_3(t)$ are solutions of the
following system of ordinary differential equations:
\begin{eqnarray*}
&& 2{c\over l^4}-{1\over 2}{\ddot l\over l}+k={1\over
2}(At+B)^2,\quad {c_3\over l_3^4}-{1\over 4}{\ddot l_3\over
l_3}=k,\\ && l\ddot v_1+2\dot l\dot v_1+4c{v_1\over
l^3}-2c_{11}{1\over l}=-2(a_1\cos\alpha+a_2\sin\alpha),\\ &&
l\ddot v_2+2\dot l\dot v_2+4c{v_2\over l^3}-2c_{12}{1\over
l}=-2(-a_1\sin\alpha+a_2\cos\alpha),\\ && l_3\ddot v_3+2\dot
l_3\dot v_3+4c_3{v_3\over l_3^3}-2c_{13}{1\over l_3}=-2a_3.
\end{eqnarray*}
Here $c,c_3,c_{11},c_{12},c_{13}$ are arbitrary real constants.

\vspace{2mm}\noindent{\bf 2. Cases of stationary magnetic
field.}\\

\noindent{\bf Case 1:}
\begin{eqnarray*}
&&e\vec H=(0,0,k),\quad k=\mbox{const}\ne 0;\\
&&eA_0=-{k^2\over 12}(x_1^2+x_2^2-2x_3^2)+ a_1x_1+a_2x_2+a_3x_3,
\end{eqnarray*}
where $\vec a=(a_1,a_2,a_3)$ is constant vector.

The coordinate system is
\[
\vec x=l {\cal O} (\vec z+\vec v).
\]
Here ${\cal O}$ is a time-dependent $3\times 3$ orthogonal matrix
${\cal O}(\alpha,\beta,\gamma)$, where $\alpha=-kt,
\beta=\mbox{const},\gamma=\mbox{const}$; $\vec z$ is one of
coordinate system, given by formulas  1-11 from (\ref{1.2});
function $l(t)$ is solution of the equation \[ k^2+{3\over
2}{\ddot l\over l}={c\over l^4}
\]
given by one of the formulas:
\[
c=\mp 1,\quad l^2=\sqrt{C_1^2\pm{1\over
k^2}}\sin\left(2\sqrt{{2\over 3}}kt\right)+C_1,
\]
for coordinate system $\vec z$ given by the formulas 1, 2, 4, 5,
6, 7, 10, 11 from the list (\ref{1.2}) and
\[
c =0, \quad l=C_1\sin\left(\sqrt{{2\over 3}}kt\right)
\]
for coordinate system $\vec z$ given by the formulas 1-11 from the
list (\ref{1.2}). Here $C_1$ is an arbitrary real constant. Vector
$\vec v$ is a solution of the following system of ordinary
differential equations:
\[
3 l \vec{\ddot v} + 6\dot l \vec{\dot v} + {2c\over l^3}\vec
v=-6{\cal O}^{-1}\vec a.
\]

\vspace{2mm}\noindent{\bf Case 2:}
\begin{eqnarray*}
&&e\vec H=(0,0,k),\quad k=\mbox{const}\ne 0;\\
&&eA_0={a\over
\sqrt{x_1^2+x_2^2+x_3^2}}-{k^2\over 12}(x_1^2+x_2^2-2x_3^2),\quad a=\mbox{const}\ne 0.
\end{eqnarray*}
The coordinate system is
\[
\vec x={\cal O}\vec z.
\]
Here ${\cal O}$ is a time-dependent $3\times 3$ orthogonal matrix
${\cal O}(\alpha,\beta,\gamma)$, where $\alpha=-kt,
\beta=\mbox{const},\gamma=\mbox{const}$ and $\vec z$ is one of the
following coordinate systems:
\begin{enumerate}
\item{spherical (formula 5 from (\ref{1.2}))},
\item{prolate spheroidal II (formula 6 from (\ref{1.2}), where one
should replace $z_3$ with $z_3=a (\coth \omega_1 \tanh \omega_2\pm
1)$)},
\item{conical (formula 11 from (\ref{1.2}))}.
\end{enumerate}

\noindent{\bf Case 3:}
\begin{eqnarray*}
&&e\vec H=(0,0,k),\quad k=\mbox{const}\ne 0;\\ &&eA_0=-{k^2\over
12}(x_1^2+x_2^2-2x_3^2)+{a_1 \over r}+a_2 {x_3\over r^3}+{a_3
\over r^2} \left({x_3\over 2r}\ln{r+x_3\over r-x_3}-1\right),
\end{eqnarray*}
where $r=\sqrt{x_1^2+x_2^2+x_3^2}$ and $a_1,a_2,a_3$ are real
constant numbers.

The coordinate system is
\[
\vec x=l {\cal O} \vec z.
\]
Here ${\cal O}$ is a time-dependent $3\times 3$ orthogonal matrix
${\cal O}(\alpha,\beta,\gamma)$, where $\alpha=-kt$,
$\beta=\gamma=0$; $\vec z$ is the spherical coordinate system,
given by formula  5 from (\ref{1.2}) and function $l(t)$ is given
by
\[
l^2=\sqrt{C_1^2\pm{1\over k^2}}\sin\left(2\sqrt{{2\over 3}}kt\right)+C_1,
\quad\mbox{or}\quad l=C_1\sin\left(\sqrt{{2\over 3}}kt\right)
\]
under condition $a_1=0$ and $l=1$ under condition $a_1\ne 0$. Here
$C_1$ is an arbitrary real constant.

\vspace{2mm}\noindent{\bf Case 4:}
\begin{eqnarray*}
&&e\vec H=(0,0,k),\quad k=\mbox{const}\ne 0;\\
&&eA_0=-{k^2\over
12}(x_1^2+x_2^2-2x_3^2)+ {a_1 \over r^+}+{a_2 \over r^-}+a_3
\left({1\over r^+}\, {\rm arctanh}{x_3^+\over r^+}- {1\over r^-}\,
{\rm arctanh}{x_3^-\over r^-}\right),
\end{eqnarray*}
where $x_3^\pm=x_3\pm a$ and $r^\pm=\sqrt{x_1^2+x_2^2+(x_3\pm
a)^2}$, and $a, a_1,a_2,a_3$ are arbitrary real constants. The
coordinate system is
\[
\vec x={\cal O}\vec z.
\]
Here ${\cal O}$ is a time-dependent $3\times 3$ orthogonal matrix
${\cal O}(\alpha,\beta,\gamma)$, where $\alpha=-kt$,
$\beta=\gamma=0$ and $\vec z$ is a prolate spheroidal coordinate
system, given by formula 6 from (\ref{1.2}).

\vspace{2mm}\noindent{\bf Case 5:}
\begin{eqnarray*}
&&e\vec H=(0,0,k),\quad k=\mbox{const}\ne 0;\\ && eA_0=-{k^2\over
12}(x_1^2+x_2^2-2x_3^2)+ 2a_1a{f_1\over f}+2a_2{x_3\over ff_1}-
2a_3 \left(a{f_1\over f}\, {\rm arccot} f_1-
  {x_3\over ff_1}\, {\rm arctanh} {x_3\over af_1}\right),
\end{eqnarray*}
where
\[
f=\sqrt{(a^2-r^2)^2+4a^2x_3^2},\quad f_1=\sqrt{{-a^2+r^2+f\over
2a^2}},\quad r=\sqrt{x_1^2+x_2^2+x_3^2},
\]
and $a, a_1,a_2,a_3$ are arbitrary real constants. The coordinate
system is
\[
\vec x={\cal O}\vec z.
\]
Here ${\cal O}$ is a time-dependent $3\times 3$ orthogonal matrix
${\cal O}(\alpha,\beta,\gamma)$, where $\alpha=-kt$,
$\beta=\gamma=0$ and $\vec z$ is an oblate spheroidal coordinate
system, given by formula 7 from (\ref{1.2}).

Note that expression for $A_0$ can be rewritten in the form
\[eA_0=-{k^2\over 12}(x_1^2+x_2^2-2x_3^2)+{a_1 +ia_2 \over \tilde
r^+}+ {a_1 -ia_2 \over \tilde r^-}+ ia_3 \left({1\over \tilde
r^+}\, {\rm arctanh}{\tilde x_3^+\over \tilde r^+}- {1\over \tilde
r^-}\, {\rm arctanh}{\tilde x_3^-\over \tilde r^-}\right),
\]
where $\tilde x_3^\pm=x_3\pm ia$ and $\tilde
r^\pm=\sqrt{x_1^2+x_2^2+(x_3\pm ia)^2}$.

\vspace{2mm}\noindent{\bf Case 6:}
\begin{eqnarray*}
&&e\vec H=(0,0,k),\quad k=\mbox{const}\ne 0;\\
&&eA_0=-{k^2\over 6}(x_1^2+x_2^2-2x_3^2) +{a_1 \over r}+a_2 x_3+{a_3 \over
r}\ln{r+x_3\over r-x_3},
\end{eqnarray*}
where $r=\sqrt{x_1^2+x_2^2+x_3^2}$ and $a_1,a_2,a_3$ are arbitrary
real constants.

The coordinate system is
\[
\vec x={\cal O}\vec z.
\]
Here ${\cal O}$ is a time-dependent $3\times 3$ orthogonal matrix
${\cal O}(\alpha,\beta,\gamma)$, where $\alpha=-kt$,
$\beta=\gamma=0$ and $\vec z$ is a parabolic coordinate system,
given by formula 8 from (\ref{1.2}).

\vspace{2mm}\noindent{\bf Case 7:}
\begin{eqnarray*}
&&e\vec H=(0,0,k),\quad k=\mbox{const}\ne 0;\\
&&eA_0=-{q\over 2}(x_1^2+x_2^2-2x_3^2) + a\ln(x_1+x_2)+a_3x_3,
\end{eqnarray*}
where $k,a,a_3$ are arbitrary real constants.

The coordinate system is
\[
x_1=e^{\omega_1}\cos(\omega_1-kt),\quad
x_2=e^{\omega_1}\sin(\omega_1-kt),\quad x_3=l_3\omega_3+v_3,
\]
where $l_3,v_3$ are solutions of the system of ordinary
differential equations
\[
{c_3\over l_3^4}-{1\over 4}{\ddot l_3\over l_3}=q,\quad l_3\ddot
v_3+2\dot l_3\dot v_3+4c_3{v_3\over l_3^3}-2c_{13}{1\over
l_3}=-2a_3. \]

Note that some of the potentials obtained have the clear physical
meaning. For instance, cases 2 and 3 under condition $k=a_2=a_3=0$
give the standard Coulomb potential. Case 4 under condition
$k=a_3=0$ gives the potential for a well-known two-center Kepler
problem, i.e., the problem of finding wave functions of electron
moving in the field of two fixed Coulomb centres with charges
$a_1, a_2$ and intercenter distance $2a$ (the model of ionized
hydrogen molecule). Coulson and Joseph \cite{cj67} showed that the
corresponding Schr\"odinger equation admits separation of
variables in the prolate coordinate system only. We obtained this
potential as a particular case of the more general potential.

\section{Concluding Remarks}

Theorem 1 provides the complete solution of the problem of
classification of the Pauli equations (\ref{1.1}), which are
solvable within the framework of the method of separation of
variables in the sense of our definition 1. According to these
theorems the coordinate systems and the vector-potentials of the
electromagnetic field $A(t, \vec x)=(A_0(t, \vec x), \vec A(t,
\vec x))$ providing separability of the corresponding Pauli
equations coincide with those for the Schr\"odinger equations with
vector-potential. So the results obtained in the article are valid
for the Schr\"odinger equation as well.

It is well-known that the possibility of variable separation in a
system of PDEs is closely connected to its symmetry properties
\cite{mill88,mill89}. Namely, solutions with separated variables
are common eigenfunctions of three matrix mutually commuting
symmetry operators of the equation under study. For all the cases
of variable separation in Pauli equation (\ref{1.1}) such matrix
second-order operators can be constructed in the explicit form, by
analogy to what has been done in \cite{zhd95a} for the
(1+2)-dimensional Schr\"odinger equation. They are expressed in
terms of the matrix coefficients of the separation equations
(\ref{2.7a})--(\ref{2.8a}).

A promising development of the research is classification and
study of superintegrable (admitting sufficiently many higher
symmetries) cases of Pauli equation. Notice that the notions of
separability and superintegrability are closely related. By now,
superintegrable physical systems can be regarded as one of the
most intensively developed and significant fields of mathematical
physics. The problem of classifying superintegrable stationary
Schr\"odinger equations with scalar potential has been solved by
Winternitz with co-workers \cite{mak67} and Evans \cite{eva90} for
space dimensions $n=2$ and $n=3$ (see also \cite{mcs00}). They
have found all potentials that allow for separability of the
corresponding Schr\"odinger equation in more than one coordinate
system. We intend to modify and generalize this approach to
$(1+3)$-dimensional Pauli equation (\ref{1.1}). A study of the
problem is in progress now and will be reported in our future
publications.

\section*{Acknowledgments}

The author thanks Renat Zhdanov for helpful suggestions and
discussions, Pavel Winternitz for important remarks and Irina
Yegorchenko for editing the manuscript.

\end{document}